\documentclass[10pt,twocolumn,twoside]{IEEEtran}
\usepackage{cite}
\ifCLASSINFOpdf
\else
  \usepackage[dvips]{graphicx}
  \graphicspath{{figures/}}
  \DeclareGraphicsExtensions{.eps}
\fi
\usepackage[cmex10,tbtags]{amsmath}
\usepackage{amssymb}
\usepackage{algorithmic}
\usepackage[caption=false,font=footnotesize]{subfig}
\usepackage{float}

\usepackage{fixltx2e}

\usepackage{psfrag}

\renewcommand{\(}{\left(}
\renewcommand{\)}{\right)}
\renewcommand{\[}{\left[}
\renewcommand{\]}{\right]}

\newcommand{\defeq}{\stackrel{\Delta}{=}}
\DeclareMathOperator{\sinc}{sinc}

\DeclareMathOperator*{\argmax}{arg\,max}

\DeclareMathOperator{\cov}{cov}

\floatstyle{ruled}
\newfloat{algorithmDSW}{tbp}{loa}
\floatname{algorithmDSW}{Algorithm}

\begin{document}
%
% paper title
% can use linebreaks \\ within to get better formatting as desired
%\title{Post-Processing with Classical Estimation to Compensate for Sampling Jitter}
\title{On the Estimation of Nonrandom Signal Coefficients from Jittered Samples}
%
%
% author names and IEEE memberships
% note positions of commas and nonbreaking spaces ( ~ ) LaTeX will not break
% a structure at a ~ so this keeps an author's name from being broken across
% two lines.
% use \thanks{} to gain access to the first footnote area
% a separate \thanks must be used for each paragraph as LaTeX2e's \thanks
% was not built to handle multiple paragraphs
%

\author{Daniel~S.~Weller*,~\IEEEmembership{Student Member,~IEEE,}
        and~Vivek~K~Goyal,~\IEEEmembership{Senior Member,~IEEE}% <-this % stops a space
\thanks{This material is based upon work supported in part by the Office of Naval Research through a National Defense Science and Engineering Graduate (NDSEG) fellowship, the National Science Foundation under CAREER Grant No.\ 0643836, Texas Instruments through its Leader University program, and Analog Devices, Inc.}% <-this % stops a space
\thanks{D. S. Weller and V. K. Goyal are with the Massachusetts Institute of Technology, Room 36-680, 77 Massachusetts Avenue, Cambridge, MA 02139 USA (phone: +1.617.324.5862; fax: +1.617.324.4290; email: \{dweller,vgoyal\}@mit.edu).}}

% note the % following the last \IEEEmembership and also \thanks - 
% these prevent an unwanted space from occurring between the last author name
% and the end of the author line. i.e., if you had this:
% 
% \author{....lastname \thanks{...} \thanks{...} }
%                     ^------------^------------^----Do not want these spaces!
%
% a space would be appended to the last name and could cause every name on that
% line to be shifted left slightly. This is one of those "LaTeX things". For
% instance, "\textbf{A} \textbf{B}" will typeset as "A B" not "AB". To get
% "AB" then you have to do: "\textbf{A}\textbf{B}"
% \thanks is no different in this regard, so shield the last } of each \thanks
% that ends a line with a % and do not let a space in before the next \thanks.
% Spaces after \IEEEmembership other than the last one are OK (and needed) as
% you are supposed to have spaces between the names. For what it is worth,
% this is a minor point as most people would not even notice if the said evil
% space somehow managed to creep in.

% The paper headers
%\markboth{IEEE Transactions on Signal Processing}{Post-Processing with Classical Estimation to Compensate for Sampling Jitter}
%\markboth{IEEE Transactions on Signal Processing}{On the Estimation of Nonrandom Signal Coefficients from Jittered Samples}
\markboth{Draft}{On the Estimation of Nonrandom Signal Coefficients from Jittered Samples}
% The only time the second header will appear is for the odd numbered pages
% after the title page when using the twoside option.
% 
% *** Note that you probably will NOT want to include the author's ***
% *** name in the headers of peer review papers.                   ***
% You can use \ifCLASSOPTIONpeerreview for conditional compilation here if
% you desire.

% If you want to put a publisher's ID mark on the page you can do it like
% this:
%\IEEEpubid{0000--0000/00\$00.00~\copyright~2007 IEEE}
% Remember, if you use this you must call \IEEEpubidadjcol in the second
% column for its text to clear the IEEEpubid mark.

% use for special paper notices
%\IEEEspecialpapernotice{(Invited Paper)}

% make the title area
\maketitle

\begin{abstract}
This paper examines the problem of estimating the parameters of a bandlimited signal from samples corrupted by random jitter (timing noise) and additive iid Gaussian noise, where the signal lies in the span of a finite basis. For the presented classical estimation problem, the Cram\'{e}r--Rao lower bound (CRB) is computed, and an Expectation-Maximization (EM) algorithm approximating the maximum likelihood (ML) estimator is developed. Simulations are performed to study the convergence properties of the EM algorithm and compare the performance both against the CRB and a basic linear estimator. These simulations demonstrate that by post-processing the jittered samples with the proposed EM algorithm, greater jitter can be tolerated, potentially reducing on-chip ADC power consumption substantially.
\end{abstract}
% IEEEtran.cls defaults to using nonbold math in the Abstract.
% This preserves the distinction between vectors and scalars. However,
% if the journal you are submitting to favors bold math in the abstract,
% then you can use LaTeX's standard command \boldmath at the very start
% of the abstract to achieve this. Many IEEE journals frown on math
% in the abstract anyway.

% Note that keywords are not normally used for peerreview papers.

% For peer review papers, you can put extra information on the cover
% page as needed:
% \ifCLASSOPTIONpeerreview
% \begin{center} \bfseries EDICS Category: DSP-RECO, SSP-PARE \end{center}
% \fi
%
% For peerreview papers, this IEEEtran command inserts a page break and
% creates the second title. It will be ignored for other modes.
\IEEEpeerreviewmaketitle

% introduction
\section{Introduction}\label{sec:intro}

\begin{figure}
\centering
\psfrag{x(t)}[][]{$x(t)$}
\psfrag{tn}[][]{$t_n$}
\psfrag{s(t)}[][]{$s(t)$}
\psfrag{wn}[][]{$w_n$}
\psfrag{ADC}[][]{ADC}
\psfrag{y}[][]{$\mathbf{y}$}
\psfrag{est}[][]{estimator}
\psfrag{offchip}[][]{off-chip}
\psfrag{out}[][]{$\mathbf{\hat{x}}$}
\includegraphics[width=3.45in]{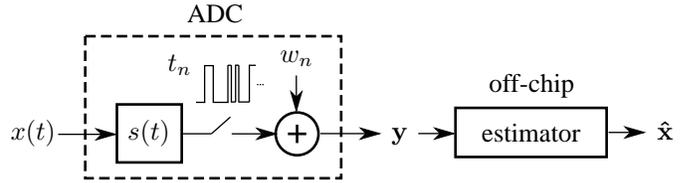}
\caption{Abstract block diagram of an ADC with off-chip post-processing. The signal $x(t)$ is filtered by the sampling prefilter $s(t)$ and sampled at time $t_n$. These samples are corrupted by additive noise $w_n$ to yield $y_n$. The post-processor estimates the parameters $\mathbf{x}$ of $x(t)$ using the vector of $N$ samples $\mathbf{y}$ from the ADC\@.\label{fig:adcblockdgrm}}
\end{figure}

An analog-to-digital converter (ADC) processes a real signal $x(t)$ to generate a sequence of observations (samples) $\{y_n\}$ at times $\{t_n\}$:
\begin{equation}
y_n = \[s(t) * x(t)\]_{t=t_n} + w_n\label{eq:intro_ADCsystem},
\end{equation}
where $s(t)$ is the sampling prefilter and $w_n$ is an additive noise term that lumps together quantization, thermal noise, and other effects. For standard sampling applications, it is assumed that the sample times are uniformly spaced by some period $T_s$ ($t_n = nT_s$, $n \in \mathbb{Z}$), where the period is small enough that the total bandwidth of $x(t)$ is less than the sampling rate $F_s = 1/T_s$. Jitter $\{z_n\}$, also known as timing noise, perturbs the sample times:
\begin{equation}
t_n = nT_s + z_n.\label{eq:intro_jittertimes}
\end{equation}
%While real ADCs may be affected by jitter,
This paper focuses on the mitigation of random jitter in a non-Bayesian estimation framework.
A simplified block diagram for the overall system is illustrated in Figure~\ref{fig:adcblockdgrm}.

The generally-accepted practice is to design clocks with low enough phase noise that the effect of jitter is negligible.
The maximum allowable jitter is set such that the effect of jitter on a sinusoid of maximum frequency and maximum amplitude is at most one-half the least significant bit level~\cite{NationalNote103}.
While making jitter negligible obviates the mitigation of jitter, this may not be possible or desirable from an overall system design perspective, because requiring jitter to have a negligible effect may mandate high power consumption in the clock circuitry.
(A model relating jitter and power is given below.)
Technological trends suggest that eventually it will be worthwhile to allow nontrivial jitter and compensate through digital post-processing:
the digital portions of mixed-signal systems like sensors and wireless transceivers continue to shrink, so the analog portions of such systems, including the ADC and its clock generator, dominate the size and power consumption of these chips.
The ability to use more power-efficient analog circuitry would enable substantial new capabilities in diverse applications like implantable medical devices and remote sensors.
One motivation of our study is to contribute to understanding the trade-off between accuracy in analog circuitry vs.\ complexity of off-chip digital post-processing of samples.

The power consumed by a typical ADC design is approximately proportional to the desired accuracy and sampling rate~\cite{Lee00}, so lower-power circuitry would produce clock signals with more jitter. Specifically, it is shown in~\cite{Uyttenhove01} that
\begin{equation}
%\frac{\text{Speed}\times\(\text{Accuracy (rms)}\)^2}{\text{Power}}\approx\text{constant}.\label{eq:intro_SAPtradeoff}
\text{Power}\propto\text{Speed}\times\(\text{Accuracy (rms)}\)^2.\label{eq:intro_SAPtradeoff}
\end{equation}
Furthermore, the analyses in~\cite{Brannon00} and~\cite{Walden99} suggest that in the large-jitter domain, every doubling of the standard deviation of the jitter reduces the effective number of bits ($\text{ENOB} = \log_2 \text{accuracy (rms)}$) by one. Thus, pre-compensating for the expected jitter in the design requires increasing the power consumption of the ADC by a factor of four for every doubling of the jitter standard deviation (e.g.\ by adding an additional level of comparators). In this paper, we instead propose block post-processing the jittered samples using classical estimation techniques off-chip. In addition to mitigating random jitter, this work may also be adapted to compensate for frequency modulated and spread-spectrum clocks, which produce lower EMI and radiation~\cite{Lee02}. Note that this block post-processing method is intended to be performed off-chip, where power consumption of an implementation of the algorithm is not important.

\subsection{Problem Formulation}

While more sophisticated signal models may be more appropriate for some applications, we concern ourselves with a signal that lies in the span of a finite basis $\{h_k(t)\}$, with $K$ basis functions. We further restrict the basis to be uniform shifts of a single smooth bandlimited function $h(t)$:
\begin{equation}
h_k(t) = h(t-kT).\label{eq:intro_hktdefined}
\end{equation}
Denote the unknown weighting parameters $x_k$; there are $K$ of them. The signal $x(t)$ then equals
\begin{equation}
x(t) = \sum_{k=0}^{K-1} x_kh(t-kT).\label{eq:intro_sigmodel}
\end{equation}
Without loss of generality, assume that $T$ equals the critical sampling period of $x(t)$, and assume this period is unity.
In this work, the signal parameters $\{x_k\}_{k=0}^{K-1}$ are unknown deterministic quantities.

While there are many possible choices of $h(t)$, when in need of a specific example in this work, we choose the sinc interpolating function $\sinc(t) = \frac{\sin(\pi t)}{\pi t}$. This basis satisfies $x(k) = x_k$, for all $k = 0,\ldots,K-1$. In general, we choose $h(t)$ to be appropriate for the class of input signals we wish to sample; we choose the sinc function because bandlimitedness is a common assumption in the context of signal processing. It is sufficient, but not necessary, in this work that $h(t)$ be analytic and bounded.

When sampling the signal $x(t)$, we will assume that the sampling prefilter is an ideal anti-aliasing filter with bandwidth $F_s$, so $[s(t) * x(t)]_{t=t_n} = x(t_n)$ for appropriately bandlimited inputs. The signal's critical sampling period is assumed to be one, but to accommodate oversampling by a factor of $M$ into our model, the ideal sample times are spaced $1/M$ time units apart. We acquire $N$ jittered samples with additive noise, $y_0,\ldots,y_{N-1}$, at this rate:
\begin{equation}
y_n = x(n/M + z_n) + w_n.\label{eq:intro_obsmodel}
\end{equation}
In this paper, we will assume that the jitter and additive noise are iid zero-mean Gaussian, with known variances equal to $\sigma_z^2$ and $\sigma_w^2$, respectively. We assume that these variances can be measured reasonably accurately through in-factory calibration, although we expect the variances to vary naturally over time due to environmental effects.

Combining the signal and observation models yields 
\begin{equation}
y_n = \sum_{k=0}^{K-1} h(n/M+z_n-k)x_k + w_n.\label{eq:intro_obssigmodel}
\end{equation}
This relationship can be expressed as a semilinear system of equations:
\begin{equation}
\mathbf{y = H(z)x+w},\label{eq:intro_obssigmodelmat}
\end{equation}
where $\mathbf{y} = [y_0,\ldots,y_{N-1}]^T$, $\mathbf{x} = [x_0,\ldots,x_{K-1}]^T$, $\mathbf{z} = [z_0,\ldots,z_{N-1}]^T$, and $\mathbf{w} = [w_0,\ldots,w_{N-1}]^T$. For notational convenience, denote the $n$th (zero-indexed) row of $\mathbf{H(z)}$ by $\mathbf{h}_n^T(z_n)$.

To keep notation compact, denote the probability density function (pdf) of $\mathbf{a}$ by $p(\mathbf{a})$, the pdf of $\mathbf{a}$ parameterized by the nonrandom vector $\mathbf{b}$ by $p(\mathbf{a;b})$, and the pdf of $\mathbf{a}$ conditioned on the random variable $\mathbf{c}$ by $p(\mathbf{a\mid c})$. The pdf is made explicit using subscripts only when necessary to avoid ambiguity. Expectations are written similarly. Also, denote the $d$-dimensional multivariate normal distribution by
%\begin{equation}
%U(\mathbf{x;a,b}) \defeq \begin{cases}\prod_{i=1}^{d} \(\frac{1}{b_i-a_i}\), & a_i \leq x_i \leq b_i,\quad i = 1,\ldots,d,\\0, & \text{otherwise},\end{cases}\label{eq:intro_udist}
%\end{equation}
%and
\begin{equation}
\mathcal{N}(\mathbf{a};\boldsymbol{\mu},\boldsymbol{\Lambda}) \defeq |2\pi\boldsymbol{\Lambda}|^{-1/2}\exp\left\{-{\textstyle\frac{1}{2}}(\mathbf{a}-\boldsymbol{\mu})^T\boldsymbol{\Lambda}^{-1}(\mathbf{a}-\boldsymbol{\mu})\right\}.\label{eq:intro_ndist}
\end{equation}

The primary objective of classical (non-Bayesian) estimation is to derive an estimator $\mathbf{\hat{x}}$ that minimizes a desired cost function $C(\mathbf{\hat{x}};\mathbf{x})$ of unknown nonrandom parameters $\mathbf{x}$. One such cost function is the mean squared error (MSE):
\begin{equation}
C(\mathbf{\hat{x}};\mathbf{x}) = \mathbb{E}_{\mathbf{y}}\[\|\mathbf{\hat{x}(y)} - \mathbf{x}\|_2^2\],\label{eq:intro_costMSE}
\end{equation}
where $\mathbb{E}_{\mathbf{y}}[\cdot]$ is the expectation with respect to $\mathbf{y}$, $\mathbf{\hat{x}(y)}$ is the estimate of the unknown parameters $\mathbf{x}$ based on the samples $\mathbf{y}$, and the observations $\mathbf{y}$ are implicitly a function of $\mathbf{x}$. However, except in certain cases, computing the estimate that minimizes this cost function, which is called the minimum MSE (MMSE) estimator, is impossible without prior knowledge about $\mathbf{x}$. Instead, it is essential to derive an estimator that relies only on the observation model and the actual observations $\mathbf{y}$. One such estimator is the maximum likelihood (ML) estimator, which maximizes the likelihood function $\ell(\mathbf{x};\mathbf{y}) \defeq p(\mathbf{y};\mathbf{x})$. The likelihood function corresponding to the signal parameter observation model in~\eqref{eq:intro_obssigmodelmat} is
\begin{equation}
\ell(\mathbf{x;y}) = \int \mathcal{N}(\mathbf{y};\mathbf{H(z)x},\sigma_w^2\mathbf{I})\mathcal{N}(\mathbf{z};\mathbf{0},\sigma_z^2\mathbf{I})\,d\mathbf{z}.\label{eq:intro_lfunc}
\end{equation}
Using the assumptions that the jitter and additive noise are iid, and the fact that the $n$th row of $\mathbf{H(z)}$ only depends on one $z_n$, the multivariate normal distributions in~\eqref{eq:intro_lfunc} are separable over $\mathbf{z}$. Thus, $p(\mathbf{y;x})$ is also separable, with $p(\mathbf{y;x}) = \prod_n p(y_n;\mathbf{x})$, and the likelihood function is the product of $N$ univariate integrals
\begin{equation}
\ell(\mathbf{x;y}) = \prod_{n=0}^{N-1}\int \mathcal{N}(y_n;\mathbf{h}_n^T(z_n)\mathbf{x},\sigma_w^2)\mathcal{N}(z_n;0,\sigma_z^2)\,dz_n.\label{eq:intro_lfuncsep}
\end{equation}
Given the likelihood function, parameters $M$, $\sigma_z^2$, and $\sigma_w^2$, MSE cost function, and observations $\mathbf{y}$, the goal of this work is use ML estimation to tolerate more jitter when estimating $\mathbf{x}$. Thus, the bulk of this paper is concerned with the evaluation of this likelihood function and the problem of maximizing it.

\subsection{Related Work}

The problem of mitigating jitter has been investigated since the early days of signal processing. The effects of jitter on the statistics of samples of a deterministic (nonrandom) bandlimited signal are briefly discussed in~\cite{Balakrishnan62}; this work also is concerned with stochastic signals and proposes an optimal linear reconstruction filter for the stochastic case. Much more work on analyzing the error and reconstructing stochastic signals from jittered samples can be found in~\cite{Brown63} and~\cite{Liu65}. However, the analysis of jittered samples of deterministic signals appears to be much more limited in the early literature.

When the sample times are irregularly spaced, but known, the problem greatly simplifies. Efficient techniques, as well as a mention of prior work, can be found in~\cite{Feichtinger95}. When the sample times are unknown, but belong to a known finite set, the jitter mitigation problem becomes a combinatorial one; \cite{Marziliano00} describes geometric and algebraic solutions to this problem of reconstructing discrete-time signals. Two block-based reconstruction methods for this finite location-set problem are described in~\cite{Tuncer07}.

However, when the set is infinitely large, or when the jitter is described by a continuous random distribution as seen here, a different approach is necessary. One contribution of this work is an Expectation-Maximization (EM) algorithm; in a similar context, \cite{Divi04} develops a similar EM algorithm for the related problem of mitigating unknown phase offsets between component ADCs in a time-interleaved ADC system. Some of the results summarized in this paper are described in greater detail in~\cite{WellerThesis}, which also provides further background material.

\subsection{Outline}

In Section~\ref{sec:background}, numerical integration using Gauss quadrature and iteration using the EM algorithm are discussed. Section~\ref{sec:CRB} presents and derives the Cram\'{e}r--Rao lower bound (CRB) on the MSE for this estimation problem. Sections~\ref{sec:linear} and~\ref{sec:ML} derive linear and ML estimators for the jitter mitigation problem; simulations comparing these estimators are discussed in Section~\ref{sec:simresults}. In conclusion, the results and contributions are summarized, and future research directions are introduced.

% background on GHquad, EM
\section{Background}\label{sec:background}

Except for certain limited choices for $h(t)$, the expression for the likelihood function in~\eqref{eq:intro_lfuncsep} has no closed form; however, various techniques exist to approximate it. One such powerful and general technique is that of quadrature, which refers to the method of approximating an integral with a finite weighted summation. The trapezoidal and Simpson's rules are elementary examples of quadrature. In particular, due to the normal distribution assumption on the jitter $z_n$, Gauss--Hermite quadrature is a natural choice of quadrature rule. Gauss--Legendre quadrature and Romberg's method are also discussed below.

Computational problems also occur when deriving the ML estimator, due to the nonconcave and high-dimensional nature of the likelihood function. One local approximation technique called the EM algorithm can be used to locate local maxima in a computationally-feasible manner. The EM algorithm is also introduced in this section.

\subsection{Numerical Integration}

Consider approximating the integral $\int f(x)w(x)\,dx$ using the summation $\sum_{j=1}^{J} w_jf(x_j)$, where $x_j$ and $w_j$ are fixed abscissas (sampling locations) and weights. This type of approximation is known generally as quadrature. When the abscissas are uniformly spaced, the summation is known as interpolatory quadrature; the trapezoidal and Simpson's rules, as well as Romberg's method~\cite{Kythe05}, are of this type. Gauss quadrature seeks greater accuracy for a given number of function evaluations by allowing the abscissas to be spaced nonuniformly. An appropriate choice of abscissas and weights (called a rule) can be precomputed for a choice of $w(x)$ and $J$ using a variety of methods, including a very efficient eigenvalue-based method derived in~\cite{Golub69}. Orthogonal polynomials satisfy a three-term recursive relationship, which is used to form a tri-diagonal matrix, whose eigenvalues are the abscissas, and whose eigenvectors yield the weights. The eigendecomposition of a tri-diagonal matrix is very efficient, so quadrature rules are very inexpensive to compute, even for very large $J$. Quadrature is particularly attractive when $f(x)$ is smooth and has bounded derivatives. This method can be applied to multivariate integration as well, although in the absence of separability, the complexity scales exponentially with the number of variables.

One weighting function of particular interest in this work is the pdf of the normal distribution. For a standard normal distribution, the associated form of quadrature is known as Gauss--Hermite quadrature, since the abscissas and weights derive from Hermite polynomials. Using elementary changes of variables, this method can be generalized to normal distributions with arbitrary mean $\mu$ and variance $\sigma^2$:
\begin{equation}
\int_{-\infty}^{\infty}f(x)\mathcal{N}(x;\mu,\sigma^2)\,dx \approx \sum_{j=1}^{J}w_jf(\sigma x_j+\mu),\label{eq:bg_GHquad}
\end{equation}
where $w_j$ and $x_j$ are the weights and abscissas for Gauss--Hermite quadrature with a standard normal weighting function. As mentioned in~\cite{Davis84}, the approximation error for Gauss--Hermite quadrature is bounded by the function's derivatives:
\begin{equation}
\left|\int_{-\infty}^{\infty}f(x)\mathcal{N}(x;\mu,\sigma^2)\,dx - \sum_{j=1}^{J}w_if(\sigma x_j + \mu)\right| \leq \frac{J!\sigma^{2J}}{(2J)!}\max_{x}\left|f^{(2J)}(x)\right|.\label{eq:bg_GHquaderror}
\end{equation}
As long as $f(x)$ is sufficiently smooth, the $(2J)!$ term in the denominator dominates the above expression for large $J$, and the approximation error goes to zero superexponentially fast. While general conditions for convergence are difficult to isolate for arbitrary $f(x)$, a sufficient condition for convergence mentioned in~\cite{Davis84} is that
\begin{equation}
\lim_{x\rightarrow\infty} \frac{|f(x)||x|^{1+\rho}}{e^{x^2}} \leq 1,\quad\text{for some}\ \rho > 0.\label{eq:bg_GHquadconverge}
\end{equation}

Many other Gauss quadrature rules exist; one simple rule also considered is called Gauss--Legendre quadrature and is defined for integrating over a finite interval $[a,b]$, with the weighting function $w(x) \equiv 1$:
\begin{equation}
\int_a^b f(x)\,dx \approx \sum_{j=1}^{J}w_jf(x_j).\label{eq:bg_GLquad}
\end{equation}
The abscissas and weights for Gauss--Legendre quadrature can be computed using the eigenvalue-based method mentioned above. Gauss--Legendre quadrature and other rules defined over a finite interval, including interpolatory quadrature methods like Simpson's rule and Romberg's method, can be extended to the infinite support case by re-mapping the variable of integration:
\begin{equation}
\int_{-\infty}^{\infty}f(x)w(x)\,dx = \int_{-\pi/2}^{\pi/2} f(\tan(y))w(\tan(y))\sec^2(y)\,dy.\label{eq:bg_tanmapping}
\end{equation}
When applied to the Gauss--Legendre quadrature rule, the new rule becomes
\begin{equation}
\int_{-\infty}^{\infty}f(x)w(x)\,dx \approx \sum_{j=1}^J w_j'f(x_j'),\label{eq:bg_GLtanquad}
\end{equation}
where $x_j' = \tan(x_j)$, and $w_j' = w(x_j')(1+(x_j')^2)w_j$.

\begin{figure}
\centering
\subfloat[][$K = 10$, $M = 4$, $\sigma_z = 0.75$, $\sigma_w = 0.1$, $J = 129$, $n = 32$.]{\includegraphics[width=3.45in]{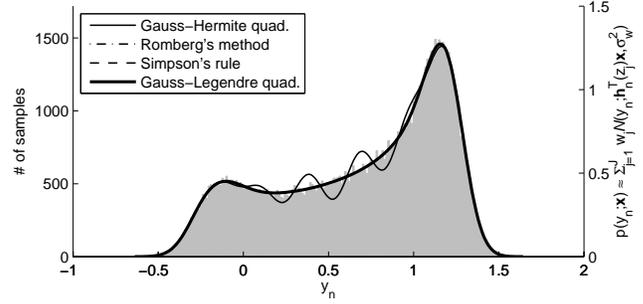}\label{fig:hermitevalid-1}}\\
\subfloat[][$K = 10$, $M = 4$, $\sigma_z = 0.01$, $\sigma_w = 0.01$, $J = 129$, $n = 34$.]{\includegraphics[width=3.45in]{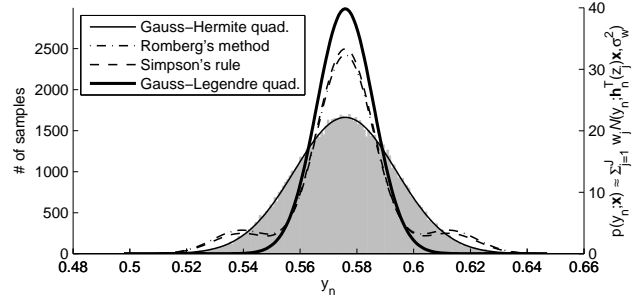}\label{fig:hermitevalid-2}}
\caption{The quadrature approximation to $p(y_n;\mathbf{x})$ is computed for a fixed, but randomly chosen, $\mathbf{x}$ on a dense grid of $y_n$ and is compared against a histogram of $100\,000$ samples $y_n$ generated using~\eqref{eq:intro_obssigmodel}. Two cases are shown to illustrate the approximation quality of~\protect\subref{fig:hermitevalid-1} Gauss-Legendre quadrature and~\protect\subref{fig:hermitevalid-2} Gauss-Hermite quadrature; the worst-case $n$ is chosen for each of these approximations. Note: the choice of $J = 129$ is used instead of $J = 100$ because Romberg's method evaluates the function $J = 2^j+1$ times for $j$ iterations.\label{fig:hermitevalid}}
\end{figure}

To compare the effectiveness of these different quadrature-based methods for numerical integration, Gauss--Hermite quadrature and Gauss--Legendre quadrature are contrasted against two more general methods, Simpson's rule and Romberg's method, by comparing each method against the marginal likelihood function $p(y_n;\mathbf{x})$, for a fixed, but randomly chosen, value of $\mathbf{x}$. The marginal likelihood function is calculated from the empirical distribution of samples generated by the observation model in~\eqref{eq:intro_obssigmodel}. As shown in Figure~\ref{fig:hermitevalid}, Gauss--Legendre quadrature approximates the likelihood function well when $\sigma_z$ is relatively large, but when $\sigma_z$ and $\sigma_w$ are both small, Gauss--Hermite quadrature is much more effective. However, other quadrature rules may be more accurate for different choices of signal basis functions $\{h_k(t)\}$.

\subsection{EM Algorithm}

The EM algorithm was introduced in~\cite{Dempster77}; a classic application of this algorithm is ML estimation in the presence of incomplete data. Consider the problem of maximizing the likelihood function $\ell(\mathbf{x};\mathbf{y})$, where $\mathbf{y}$ depends on some latent random variables $\mathbf{z}$. The observations $\mathbf{y}$ are described as the incomplete data. We augment this incomplete data with some subset of latent (hidden) variables to form the complete data. The underlying assumption of the EM algorithm is that knowledge of the complete data makes the ML estimation problem easier to solve.

The EM algorithm consists of repeatedly maximizing the function
\begin{equation}
Q\(\mathbf{x};\mathbf{\hat{x}}^{(i-1)}\) = \mathbb{E}\[\log p(\mathbf{y,z;x})\mid\mathbf{y};\mathbf{\hat{x}}^{(i-1)}\]\label{eq:bg_EMstep}
\end{equation}
with respect to the desired parameters $\mathbf{x}$; the maximizing value becomes $\mathbf{\hat{x}}^{(i)}$, which is used in the next iteration. As long as the original likelihood function is bounded above, and some other mild conditions are satisfied, this algorithm is guaranteed to converge to a local maximum of the likelihood function~\cite{Dempster77}.

Much has been written about the convergence rate of EM algorithms. In~\cite{Herzet07}, the rate of convergence of the EM algorithm is related to the difference in the CRB using the incomplete data and the CRB using the complete data (incomplete data + latent variables). The supplemented EM algorithm in~\cite{Meng91} also obtains Fisher information estimates, conditioned on the observations $\mathbf{y}$, which can be used to evaluate the quality of the resulting approximation to the ML estimate.

Since the likelihood function in~\eqref{eq:intro_lfunc} is not in general strictly concave, the presence of many critical points is a potential problem for any local algorithm. Simulated annealing~\cite{Kirkpatrick83} and other methods can be combined with the EM algorithm to improve robustness to getting trapped in local extrema.

\section{Approximating the CRB}\label{sec:CRB}

Minimizing the MSE without access to prior information about the unknown parameters $\mathbf{x}$ may be impossible in the general case. However, even in situations when the MMSE estimator cannot be computed, the Cram\'{e}r--Rao lower bound on the minimum achievable MSE by an unbiased estimator may be straightforward to compute. The $\text{CRB}_\mathbf{y}(\mathbf{x})$, where $\mathbf{x}$ is the variable to estimate from observations $\mathbf{y}$, is defined to be the trace of the inverse of the Fisher information matrix $\mathbf{I_y}(\mathbf{x})$, which is defined as
\begin{equation}
\mathbf{I_{y}}(\mathbf{x}) \defeq \mathbb{E}\[\(\frac{\partial\log\ell(\mathbf{x;y})}{\partial \mathbf{x}}\)\(\frac{\partial\log\ell(\mathbf{x;y})}{\partial \mathbf{x}}\)^T\].\label{eq:CRB_fisher}
\end{equation}
Assuming the likelihood function satisfies the regularity condition
\begin{equation}
\mathbb{E}\[\frac{1}{\ell(\mathbf{x;y})}\frac{\partial^2\ell(\mathbf{x;y})}{\partial\mathbf{x}\partial\mathbf{x}^T}\] = \int_{-\infty}^{\infty} \frac{\partial^2\ell(\mathbf{x;y})}{\partial\mathbf{x}\partial\mathbf{x}^T}\,d\mathbf{y} = \mathbf{0},\label{eq:CRB_fisherregcon}
\end{equation}
the Fisher information matrix can be expressed in terms of the Hessian of the log-likelihood:
\begin{align}
[\mathbf{I_{y}}(\mathbf{x})]_{j,k} &= \mathbb{E}\[\frac{\partial\log\ell(\mathbf{x;y})}{\partial x_j}\frac{\partial\log\ell(\mathbf{x;y})}{\partial x_k}\]\\
&= \mathbb{E}\[\frac{1}{\ell(\mathbf{x;y})^2}\frac{\partial\ell(\mathbf{x;y})}{\partial x_j}\frac{\partial\ell(\mathbf{x;y})}{\partial x_k}\]\\
&= \mathbb{E}\[\frac{1}{\ell(\mathbf{x;y})^2}\frac{\partial\ell(\mathbf{x;y})}{\partial x_j}\frac{\partial\ell(\mathbf{x;y})}{\partial x_k} - \frac{1}{\ell(\mathbf{x;y})}\frac{\partial^2\ell(\mathbf{x;y})}{\partial x_j \partial x_k}\]\\
&= \mathbb{E}\[\frac{\partial}{\partial x_k}\(\frac{-1}{\ell(\mathbf{x;y})}\frac{\partial\ell(\mathbf{x;y})}{\partial x_j}\)\]\\
\mathbf{I_{y}}(\mathbf{x}) &= -\mathbb{E}\[\frac{\partial^2\log\ell(\mathbf{x;y})}{\partial\mathbf{x}\partial\mathbf{x}^T}\].\label{eq:CRB_fisher2}
\end{align}
A sufficient condition for the regularity condition in~\eqref{eq:CRB_fisherregcon} is that differentiation and integration interchange, since
\begin{equation}
\int_{-\infty}^{\infty} \frac{\partial^2\ell(\mathbf{x;y})}{\partial\mathbf{x}\partial\mathbf{x}^T}\,d\mathbf{y} = \frac{\partial^2}{\partial\mathbf{x}\partial\mathbf{x}^T}\[\int_{-\infty}^{\infty} \ell(\mathbf{x;y})\,d\mathbf{y}\] = \frac{\partial^2}{\partial\mathbf{x}\partial\mathbf{x}^T}(1) = 0.\label{eq:CRB_fisherregich}
\end{equation}
By basic analysis, uniform convergence of the integral in likelihood function in~\eqref{eq:intro_lfunc} implies that the above regularity condition holds.

Since the likelihood function is separable, the log-likelihood function can be expressed as the summation of marginal log-likelihood functions; i.e.
\begin{equation}
\log\ell(\mathbf{x;y}) = \sum_{n=0}^{N-1}\log p(y_n;\mathbf{x}),\label{eq:CRB_llsep}
\end{equation}
which means that~\eqref{eq:CRB_fisher2} can be rewritten as (again, assuming regularity conditions are satisfied)
\begin{equation}
\mathbf{I_{y}}(\mathbf{x}) = \sum_{n=0}^{N-1} \mathbb{E}\[\(\frac{\partial\log p(y_n;\mathbf{x})}{\partial\mathbf{x}}\)\(\frac{\partial\log p(y_n;\mathbf{x})}{\partial\mathbf{x}}\)^T\].\label{eq:CRB_fishersep}
\end{equation}

The marginal pdf $p(y_n;\mathbf{x})$ can be computed numerically using quadrature:
\begin{equation}
p(y_n;\mathbf{x}) \approx \sum_{j=1}^J w_j\mathcal{N}(y_n;\mathbf{h}_n^T(z_j)\mathbf{x},\sigma_w^2),\label{eq:CRB_pynGHapprox}
\end{equation}
where $z_j$ and $w_j$ are the abscissas and weights for the chosen quadrature rule. As depicted in Figure~\ref{fig:hermitevalid}, Gauss--Hermite quadrature is a good choice for small $\sigma_z$, and Gauss--Legendre quadrature is more accurate for larger $\sigma_z$ ($> 0.1$). For all simulations in this paper, we use $J = 100$ unless otherwise specified.

The derivative of the marginal pdf can be approximated similarly:
\begin{equation}
\begin{split}
\frac{\partial p(y_n;\mathbf{x})}{\partial\mathbf{x}} &= \int \frac{(y_n-\mathbf{h}_n^T(z_n)\mathbf{x})\mathbf{h}_n(z_n)}{\sigma_w^2}\mathcal{N}(y_n;\mathbf{h}_n^T(z_n)\mathbf{x},\sigma_w^2)\mathcal{N}(z_n;0,\sigma_z^2)\,dz_n\\
&\approx \sum_{j=1}^J w_j\frac{(y_n-\mathbf{h}_n^T(z_j)\mathbf{x})\mathbf{h}_n(z_j)}{\sigma_w^2}\mathcal{N}(y_n;\mathbf{h}_n^T(z_j)\mathbf{x},\sigma_w^2).\end{split}\label{eq:CRB_dpynGHapprox}
\end{equation}
Since $\frac{\partial}{\partial\mathbf{x}}[\log p(y_n;\mathbf{x})]$ is equal to $\frac{1}{p(y_n;\mathbf{x})}\frac{\partial p(y_n;\mathbf{x})}{\partial\mathbf{x}}$, combining~\eqref{eq:CRB_pynGHapprox} and~\eqref{eq:CRB_dpynGHapprox} yields a complicated approximation to the expression inside the expectation in~\eqref{eq:CRB_fishersep}:
\begin{multline}
\(\frac{\partial\log p(y_{n};\mathbf{x})}{\partial\mathbf{x}}\)\(\frac{\partial\log p(y_{n};\mathbf{x})}{\partial\mathbf{x}}\)^T\\
\approx \(\frac{\sum_{j=1}^J w_j(y_n-\mathbf{h}_n^T(z_j)\mathbf{x})\mathbf{h}_n(z_j)\mathcal{N}(y_n;\mathbf{h}_n^T(z_j)\mathbf{x},\sigma_w^2)}{\sigma_{w}^{2}\sum_{j=1}^J w_j\mathcal{N}(y_n;\mathbf{h}_n^T(z_j)\mathbf{x},\sigma_w^2)}\)\\
\cdot\(\frac{\sum_{j=1}^J w_j(y_n-\mathbf{h}_n^T(z_j)\mathbf{x})\mathbf{h}_n(z_j)\mathcal{N}(y_n;\mathbf{h}_n^T(z_j)\mathbf{x},\sigma_w^2)}{\sigma_{w}^{2}\sum_{j=1}^J w_j\mathcal{N}(y_n;\mathbf{h}_n^T(z_j)\mathbf{x},\sigma_w^2)}\)^{T}.\label{eq:CRB_fisherexpGHapprox}
\end{multline}
For convenience, denote the above approximation $\mathbf{F}_n(y_n;\mathbf{x})$.

Now, consider computing the Fisher information matrix from this approximation:
\begin{equation}
\mathbf{I_{y}}(\mathbf{x}) \approx \sum_{n=0}^{N-1} \mathbb{E}\[\mathbf{F}_n(y_n;\mathbf{x})\].\label{eq:CRB_fishersepGHapprox}
\end{equation}
To compute this expectation, a numerical method is needed again. The expectation is with respect to the distribution $p(y_n;\mathbf{x})$, which is approximated in~\eqref{eq:CRB_pynGHapprox} with a Gaussian mixture, so Monte Carlo sampling is a convenient method to approximate this expectation. Generating $S$ samples $y_{n,s}$ from the Gaussian mixture $\sum_{j=1}^J w_j\mathcal{N}(y_n;\mathbf{h}_n^T(z_j)\mathbf{x},\sigma_w^2)$ and averaging the corresponding function values $\mathbf{F}_n(y_{n,s};\mathbf{x})$, the Fisher information matrix can be computed as
\begin{equation}
\mathbf{I_y(x)} \approx \frac{1}{S}\sum_{n=0}^{N-1}\sum_{s=1}^{S} \mathbf{F}_n(y_{n,s};\mathbf{x}).\label{eq:CRB_fisherMCapprox}
\end{equation}
Once this matrix is computed and inverted, the trace gives the Cram\'{e}r--Rao lower bound for that choice of parameter $\mathbf{x}$. Due to matrix inversion, to ensure an accurate CRB estimate, we use $J = 1000$ for the quadratures in~\eqref{eq:CRB_fisherexpGHapprox}.

How much does the CRB decrease when $\mathbf{z}$ is assumed given? Comparing the $\text{CRB}_{\mathbf{y}}(\mathbf{x})$ against the $\text{CRB}_{\mathbf{y,z}}(\mathbf{x})$ of the jitter-augmented data will be important later when analyzing the EM algorithm design. The Fisher information matrix in this case is equal to
\begin{equation}
\mathbf{I_{y,z}(x)} = -\sum_{n=0}^{N-1}\mathbb{E}\[\frac{\partial^2 \log p(y_n,z_n;\mathbf{x})}{\partial\mathbf{x}\partial\mathbf{x}^T}\].\label{eq:CRB_Iyzsep}
\end{equation}
Of course, since
\begin{equation}
-\log p(y_n,z_n;\mathbf{x}) = \frac{1}{2\sigma_w^2}(y_n-\mathbf{h}_n^T(z_n)\mathbf{x})^2 + \frac{1}{2\sigma_z^2}z_n^2 + \text{constant},\label{eq:CRB_lpyzsep}
\end{equation}
and the Hessian matrix with respect to $\mathbf{x}$ is
\begin{equation}
\frac{\partial^2}{\partial\mathbf{x}\partial\mathbf{x}^{T}}\log p(y_n,z_n;\mathbf{x}) = -\frac{1}{\sigma_w^2}\mathbf{h}_n(z_n)\mathbf{h}_n^T(z_n),\label{eq:CRB_ddlpyzsep}
\end{equation}
the jitter-augmented Fisher information matrix is
\begin{equation}
\mathbf{I_{y,z}(x)} = \frac{1}{\sigma_w^2} \sum_{n=0}^{N-1} \mathbb{E}\[\mathbf{h}_n(z_n)\mathbf{h}_n^T(z_n)\],\label{eq:CRB_Iyzsep2}
\end{equation}
where the expectation can be approximated numerically using quadrature or Monte Carlo approximation. The jitter-augmented $\text{CRB}_{\mathbf{y,z}}(\mathbf{x})$ is the trace of the inverse of this matrix. We will return to the question of the difference of the two CRBs later in Section~\ref{sec:simresults}, after we discuss ML estimation using an EM algorithm.

% linear estimator
\section{Linear Estimation}\label{sec:linear}

In this paper, an estimator is said to be linear if it is a linear function of the observations; such an estimator has the form
\begin{equation}
\mathbf{\hat{x}}_{\text{L}}(\mathbf{y}) = \mathbf{Ay},\label{eq:lin_linform}
\end{equation}
where the matrix $\mathbf{A}$ is fixed.\footnote{Sometimes, affine estimators $\mathbf{\hat{x}(y) = Ay+b}$ are considered to be linear as well. However, as we will concern ourselves with unbiased estimators, $\mathbf{b}$ would turn out to be necessarily zero.}

For the semilinear observation model in~\eqref{eq:intro_obssigmodelmat}, a linear estimator is unbiased if and only if $\mathbf{A}\mathbb{E}[\mathbf{H(z)}] = \mathbf{I}$. Since $\mathbb{E}[\mathbf{H(z)}]$ is a tall matrix, assuming it has full column rank, one possible linear unbiased estimator is
\begin{equation}
\mathbf{\hat{x}}_{\text{L}}(\mathbf{y}) = \mathbb{E}[\mathbf{H(z)}]^{\dag}\mathbf{y},\label{eq:lin_linestpinv}
\end{equation}
where $\mathbb{E}[\mathbf{H(z)}]^{\dag} \defeq \(\mathbb{E}[\mathbf{H(z)}]^{T}\mathbb{E}[\mathbf{H(z)}]\)^{-1}\mathbb{E}[\mathbf{H(z)}]^{T}$ is the left pseudoinverse. This estimator is only one such linear unbiased estimator; more generally, any matrix that lies in the nullspace of $\mathbb{E}[\mathbf{H(z)}]$ can be added to the pseudoinverse and yield an unbiased estimator.

The question then remains of how to obtain the best linear unbiased estimator (BLUE), in the MMSE sense. In the context of a simple linear observation model $\mathbf{y = Hx+w}$, with Gaussian noise $w$, the BLUE is elementary to find (see~\cite{Kay93}), and it is also the ML and efficient minimum variance unbiased estimator (MVUE). If we choose $\mathbf{z = 0}$ to be deterministic (no jitter) in the observation model, the corresponding BLUE/efficient estimator would be
\begin{equation}
\mathbf{\hat{x}}_{\text{eff}|\mathbf{z=0}}(\mathbf{y}) = \(\mathbf{H(0)}^T\mathbf{H(0)}\)^{-1}\mathbf{H(0)}^T\mathbf{y}.\label{eq:lin_blueestnoz}
\end{equation}
The performance of this estimator when applied to the proper (jittered) observation model will be used as one baseline for MSE improvement for the proposed estimators.

As derived previously in~\cite{WellerThesis}, the BLUE for the semilinear model~\eqref{eq:intro_obssigmodelmat} is
\begin{equation}
\mathbf{\hat{x}}_{\text{BLUE}}(\mathbf{y}) = \(\mathbb{E}[\mathbf{H(z)}]^T\boldsymbol{\Lambda}_{\mathbf{y}}^{-1}\mathbb{E}[\mathbf{H(z)}]\)^{-1}\mathbb{E}[\mathbf{H(z)}]^T\boldsymbol{\Lambda}_{\mathbf{y}}^{-1}\mathbf{y},\label{eq:lin_blueest}
\end{equation}
where the covariance matrix of the data $\boldsymbol{\Lambda}_{\mathbf{y}}$ depends on the value of the parameters:
\begin{equation}
\boldsymbol{\Lambda}_{\mathbf{y}} = \mathbb{E}[\mathbf{H(z)x}\mathbf{x}^T\mathbf{H(z)}^T] - \mathbb{E}[\mathbf{H(z)}]\mathbf{x}\mathbf{x^T}\mathbb{E}[\mathbf{H(z)}]^T + \sigma_w^2\mathbf{I}.\label{eq:lin_blueestcov}
\end{equation}
The BLUE estimator, in general, is not a valid estimator, since it depends on the true value of the unknown $\mathbf{x}$. Two sufficient conditions for the estimator to be valid are: $\boldsymbol{\Lambda}_{\mathbf{y}}$ is a scalar matrix, in which case, the covariance matrix commutes across multiplication, or $\boldsymbol{\Lambda}_{\mathbf{y}}$ does not depend on $\mathbf{x}$. Since $z_m$ and $z_n$ are independent for $m \neq n$, off-diagonal elements of $\boldsymbol{\Lambda}_{\mathbf{y}}$ are zero. For the covariance matrix to be a scalar matrix that commutes over matrix multiplication for all $\mathbf{x}$, $\cov(\mathbf{h}_n(z_n))$ must be equal for all $n$. However, this equality generally does not hold due to oversampling. Also, the covariance matrix clearly depends on $\mathbf{x}$ when $\cov(\mathbf{h}_n(z_n))$ is nonzero for some $n$. When the covariance matrix $\boldsymbol{\Lambda}_{\mathbf{y}}$ is a scalar matrix, the BLUE estimator is equal to $\mathbf{\hat{x}}_{\text{L}}(\mathbf{y})$.

\begin{figure}
\centering
\includegraphics[width=3.45in]{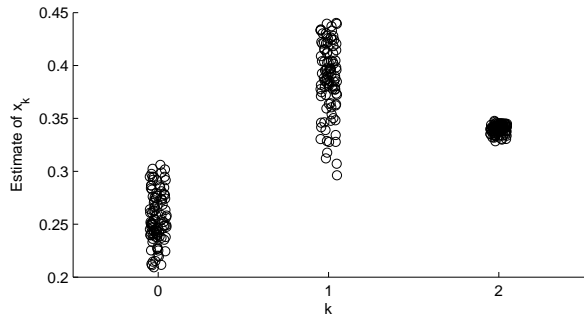}
\caption{The best linear unbiased estimator (BLUE) is computed for different choices of $\mathbf{x}$ (holding $\mathbf{y}$ fixed), using~\eqref{eq:lin_blueest}. In this example, $K = 3$, $M = 2$, and $\sigma_z = \sigma_w = 0.25$.\label{fig:BLUEvalid}}
\end{figure}

To conclusively demonstrate that the BLUE is not a valid estimator, the estimator is computed for a fixed value of $\mathbf{y}$ and varying $\mathbf{x}$; the results are shown in Figure~\ref{fig:BLUEvalid}. Clearly, since the estimates of $x_k$ vary depending on the value of $\mathbf{x}$ used in~\eqref{eq:lin_blueest}, the estimator is not valid. Thus, an MSE-optimal linear estimator does not exist for this problem, and we will utilize the estimator in~\eqref{eq:lin_linestpinv}.

% ML estimators
\section{ML Estimation}\label{sec:ML}

Given a semilinear model as in~\eqref{eq:intro_obssigmodelmat}, we would not expect the optimal MMSE estimator to have a linear form. Indeed, as shown in the previous section, for most signal models and priors on the jitter, the BLUE does not even exist. To improve upon linear estimation, and reduce the MSE, we move to maximum likelihood estimation.

Consider the problem of maximizing the likelihood function in~\eqref{eq:intro_lfuncsep}; since the logarithm is an increasing function, we can perform the optimization by maximizing the log-likelihood:
\begin{equation}
\mathbf{\hat{x}}_{\text{ML}}(\mathbf{y}) = \argmax_{\mathbf{x}} \sum_{n=0}^{N-1} \log p(y_n;\mathbf{x}).\label{eq:ML_mlestform}
\end{equation}
However, since the marginal pdf does not have a closed form, and neither do its derivatives, performing the necessary optimization is difficult. Numerical techniques may be applied directly to~\eqref{eq:ML_mlestform}, and various general-purpose methods have been studied extensively throughout the literature. An iterative joint maximization method proposed in~\cite{Kusuma08} attempts to approximate the ML estimate by alternating between maximizing $p(\mathbf{z}\mid\mathbf{y;x})$ with respect to $\mathbf{z}$ and $p(\mathbf{y}\mid\mathbf{z;x})$ with respect to $\mathbf{x}$. One method that explicitly takes advantage of the special structure in~\eqref{eq:ML_mlestform} is the EM algorithm.

\subsection{ML Estimation using the EM Algorithm}

Consider the function $Q\(\mathbf{x};\mathbf{\hat{x}}^{(i-1)}\)$ in~\eqref{eq:bg_EMstep}. The expression for $\log p(y_n,z_n;\mathbf{x})$ is in~\eqref{eq:CRB_lpyzsep}, and summing these together (without the minus sign) gives
\begin{equation}
\log p(\mathbf{y,z;x}) = -\frac{1}{2\sigma_w^2} \|\mathbf{y-H(z)x}\|_2^2 - \frac{1}{2\sigma_z^2}\|\mathbf{z}\|_2^2 + \text{constants}.\label{eq:ML_lpyz}
\end{equation}
Expanding and substituting into the expectation in~\eqref{eq:bg_EMstep} yields
\begin{equation}
\begin{split}
Q(\mathbf{x};\mathbf{\hat{x}}^{(i-1)}) &= \frac{-1}{2\sigma_{w}^{2}}\(\mathbf{y}^{T}\mathbf{y}-2\mathbf{y}^{T}\mathbb{E}\[\mathbf{H}(\mathbf{z})\mid \mathbf{y};\mathbf{\hat{x}}^{(i-1)}\]\mathbf{x} + \mathbf{x}^{T}\mathbb{E}\[\mathbf{H}(\mathbf{z})^{T}\mathbf{H}(\mathbf{z})\mid \mathbf{y};\mathbf{\hat{x}}^{(i-1)}\]\mathbf{x}\)\\
&\quad- \frac{1}{2\sigma_{z}^{2}}\mathbb{E}\[\mathbf{z}^{T}\mathbf{z}\mid \mathbf{y};\mathbf{\hat{x}}^{(i-1)}\] + \text{constants}.\end{split}\label{eq:ML_Qexp}
\end{equation}

We want to find the value of $\mathbf{x}$ that maximizes this expression. Noticing that~\eqref{eq:ML_Qexp} is quadratic in $\mathbf{x}$, the candidate value $\mathbf{x}$ satisfies the linear system
\begin{equation}
\mathbb{E}\[\mathbf{H(z)}^T\mathbf{H(z)}\mid\mathbf{y};\mathbf{\hat{x}}^{(i-1)}\]\mathbf{x} = \mathbb{E}\[\mathbf{H(z)}\mid\mathbf{y};\mathbf{\hat{x}}^{(i-1)}\]^T\mathbf{y}.\label{eq:ML_Qmax}
\end{equation}
Also, the Hessian matrix is negative-definite, so~\eqref{eq:ML_Qexp} is strictly concave, and the candidate point $\mathbf{x}$ is the unique maximum $\mathbf{\hat{x}}^{(i)}$. All that remains to specify the EM algorithm is to approximate the expectations in~\eqref{eq:ML_Qmax}.

Using Bayes' rule and the separability of both $p(\mathbf{y,z;x})$ and $p(\mathbf{y;x})$, the posterior distribution of the jitter is also separable:
\begin{align}
p\(\mathbf{z}\mid\mathbf{y};\mathbf{\hat{x}}^{(i-1)}\) &= \prod_{n=0}^{N-1}\frac{p\(y_n\mid z_n;\mathbf{\hat{x}}^{(i-1)}\)p(z_n)}{p\(y_n;\mathbf{\hat{x}}^{(i-1)}\)}\label{eq:ML_pzyx}\\
&= \prod_{n=0}^{N-1} p\(z_n\mid y_n;\mathbf{\hat{x}}^{(i-1)}\).\label{eq:ML_pzyxsep}
\end{align}
Thus, the expectations are also separable into univariate expectations:
\begin{align}
\mathbb{E}\[\mathbf{H(z)}^T\mathbf{H(z)}\mid\mathbf{y};\mathbf{\hat{x}}^{(i-1)}\] &= \sum_{n=0}^{N-1} \mathbb{E}\[\mathbf{h}_n(z_n)\mathbf{h}_n^T(z_n)\mid y_n;\mathbf{\hat{x}}^{(i-1)}\];\label{eq:ML_EHHyxexp}\\
\mathbb{E}\[\mathbf{H(z)}\mid\mathbf{y};\mathbf{\hat{x}}^{(i-1)}\]_{n,:} &= \mathbb{E}\[\mathbf{h}_n^T(z_n)\mid y_n;\mathbf{\hat{x}}^{(i-1)}\].\label{eq:ML_EHyxexp}
\end{align}
The subscript after the left-side expectation in~\eqref{eq:ML_EHyxexp} denotes the $n$th (zero-indexed) row of the matrix. The distribution $p\(y_n;\mathbf{\hat{x}}^{(i-1)}\)$ is constant with respect to $z_n$, and can be evaluated using quadrature, as in~\eqref{eq:CRB_pynGHapprox}. Approximating each of the univariate expectations in~\eqref{eq:ML_EHHyxexp} and~\eqref{eq:ML_EHyxexp} with quadrature yields
\begin{align}
\mathbb{E}\[\mathbf{H(z)}^T\mathbf{H(z)}\mid\mathbf{y};\mathbf{\hat{x}}^{(i-1)}\] &\approx \sum_{n=0}^{N-1} \frac{1}{p\(y_n;\mathbf{\hat{x}}^{(i-1)}\)}\sum_{j=1}^{J}w_j\mathbf{h}_n(z_j)\mathbf{h}_n^T(z_j)p\(y_n \mid z_j;\mathbf{\hat{x}}^{(i-1)}\),\ \text{and}\label{eq:ML_GHEHHyxexp}\\
\mathbb{E}\[\mathbf{H(z)}\mid\mathbf{y};\mathbf{\hat{x}}^{(i-1)}\]_{n,:} &\approx \frac{1}{p\(y_n;\mathbf{\hat{x}}^{(i-1)}\)}\sum_{j=1}^{J}w_j\mathbf{h}_n^T(z_j)p\(y_n \mid z_j;\mathbf{\hat{x}}^{(i-1)}\).\label{eq:ML_GHEHyxexp}
\end{align}

The complexity of each iteration of this algorithm appears to be linear in the number of samples, although the rate of convergence (and thus, the number of iterations required) may also vary with the
number of samples, or other factors. The convergence rate, as well as susceptibility to initial conditions (since the EM algorithm only guarantees local convergence), are the subject of simulations in this work and in~\cite{WellerThesis}.

The EM algorithm for ML estimation is summarized in Algorithm~\ref{alg:ml_emalg}.
\begin{algorithmDSW}
\begin{algorithmic}
\REQUIRE $p(\mathbf{y,z;x})$, $\mathbf{y}$, $\mathbf{\hat{x}}^{(0)}$, $I$, $J$, $\delta$, $\epsilon$
\STATE $i \leftarrow 0$
\STATE Compute $J$-term quadrature rule (using e.g., the eigendecomposition method in~\cite{Golub69}) for use in below approximations (use $J = 100$).
\REPEAT
\STATE $i \leftarrow i+1$
\FOR{$n = 0$ to $N-1$}
\STATE Approximate $p(y_n;\mathbf{\hat{x}}^{(i-1)})$ using~\eqref{eq:CRB_pynGHapprox}.
\STATE Compute $\mathbb{E}\[\mathbf{h}_n(z_n)\mathbf{h}_n^T(z_n)\mid y_n;\mathbf{\hat{x}}^{(i-1)}\]$ using~\eqref{eq:ML_GHEHHyxexp}.
\STATE Approximate $\mathbb{E}\[\mathbf{H(z)}\mid\mathbf{y};\mathbf{\hat{x}}^{(i-1)}\]_{n,:}$ using~\eqref{eq:ML_GHEHyxexp}.
\ENDFOR
\STATE Solve for $\mathbf{\hat{x}}^{(i)}$ using~\eqref{eq:ML_Qmax} and the above approximations.
\UNTIL{$i = I$ \textbf{or} $\|\mathbf{\hat{x}}^{(i)} - \mathbf{\hat{x}}^{(i-1)}\|_2 < \delta$ \textbf{or} $|\log\ell(\mathbf{\hat{x}}^{(i)};\mathbf{y}) - \log\ell(\mathbf{\hat{x}}^{(i-1)};\mathbf{y})| < \epsilon$}
\RETURN $\mathbf{\hat{x}}^{(i)}$
\end{algorithmic}
\caption{Pseudocode for the EM algorithm for computing ML estimates for the unknown signal parameters $\mathbf{x}$.\label{alg:ml_emalg}}
\end{algorithmDSW}

% simulations, results
\section{Simulation Results}\label{sec:simresults}

The objectives of the simulations presented here are to (a) analyze the behavior of the proposed EM algorithm for approximating the ML estimator, and to (b) compare the performance of this estimator to both the Cram\'{e}r--Rao bound and that of linear parameter estimation. The convergence behavior is studied in detail in~\cite{WellerThesis} for periodic bandlimited signals. In this work, experiments to determine convergence behavior and sensitivity to initial conditions are conducted for the sinc basis signal model described in Section~\ref{sec:intro}. In all experiments, we utilize MATLAB to generate signals with pseudo-random parameters and noise and apply the algorithms described to the samples of these signals. For a factor of $M$ oversampling, we generate $N = K \, M$ samples.

\subsection{Convergence Analysis}

\begin{figure}
\centering
\subfloat[][$K = 10$, $\sigma_z = 0.25$, $\sigma_w = 0.1$, $M$ varies.]{\includegraphics[width=3.45in]{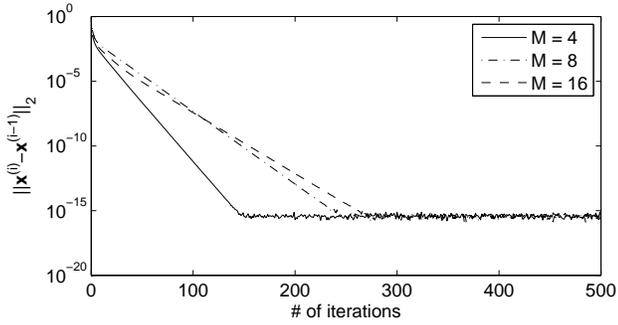}\label{fig:converge-1}}\\
\subfloat[][$K = 10$, $M = 4$, $\sigma_w = 0.1$, $\sigma_z$ varies.]{\includegraphics[width=3.45in]{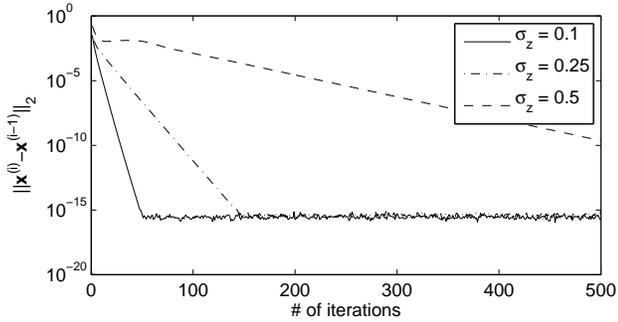}\label{fig:converge-2}}\\
\subfloat[][$K = 10$, $M = 4$, $\sigma_z = 0.25$, $\sigma_w$ varies.]{\includegraphics[width=3.45in]{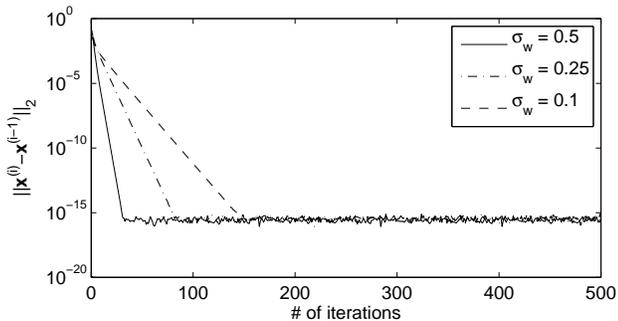}\label{fig:converge-3}}\\
\caption{The convergence of the EM algorithm depends on the choice of parameters $M$, $\sigma_z$, and $\sigma_w$, as demonstrated in the above plots.\label{fig:converge}}
\end{figure}

While guaranteed to converge, the EM algorithm would be of little use if it did not converge quickly. The rate of convergence of the EM algorithm is studied for several choices of $M$, $\sigma_z$, and $\sigma_w$, and trends are presented in Figure~\ref{fig:converge}. The rate of convergence is exponential, and the rate decreases with increasing $M$, increasing $\sigma_z$, and decreasing $\sigma_w$.

As mentioned in Section~\ref{sec:background}, the rate of convergence of the EM algorithm is related to the difference between the CRBs of the complete and the incomplete data. As shown later in Figure~\ref{fig:crb}, the difference between the CRBs for the complete data and incomplete data increases exponentially with $\sigma_z$. This relationship coincides with the convergence behavior observed in Figure~\ref{fig:converge-2}. Although these experiments evaluated $500$ iterations of the EM algorithm, the results suggest that $100$ iterations would suffice as long as the jitter standard deviation $\sigma_z$ is not too large. Also, $10^{-8}$ is chosen as a reasonable stopping criterion for change in $\mathbf{x}^{(i)}$ and change in log-likelihood between iterations ($\delta$ and $\epsilon$ in Algorithm~\ref{alg:ml_emalg}, respectively).

\subsection{Sensitivity to Initial Conditions}

\begin{figure}
\centering
\subfloat[][$K = 10$, $\sigma_z = 0.25$, $\sigma_w = 0.25$, and varying $M$.]{\includegraphics[width=3.45in]{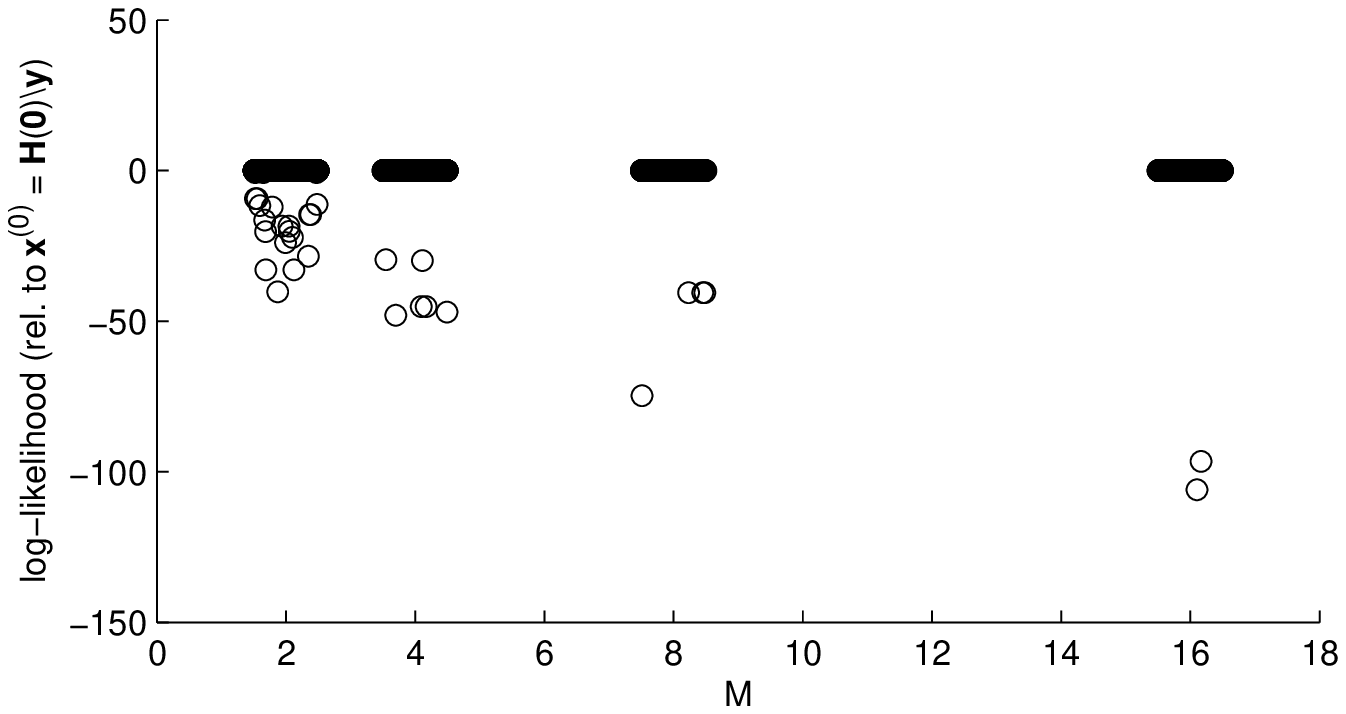}\label{fig:initscat-Ms}}\\
\subfloat[][$K = 10$, $M = 8$, $\sigma_w = 0.25$, and varying $\sigma_z$.]{\includegraphics[width=3.45in]{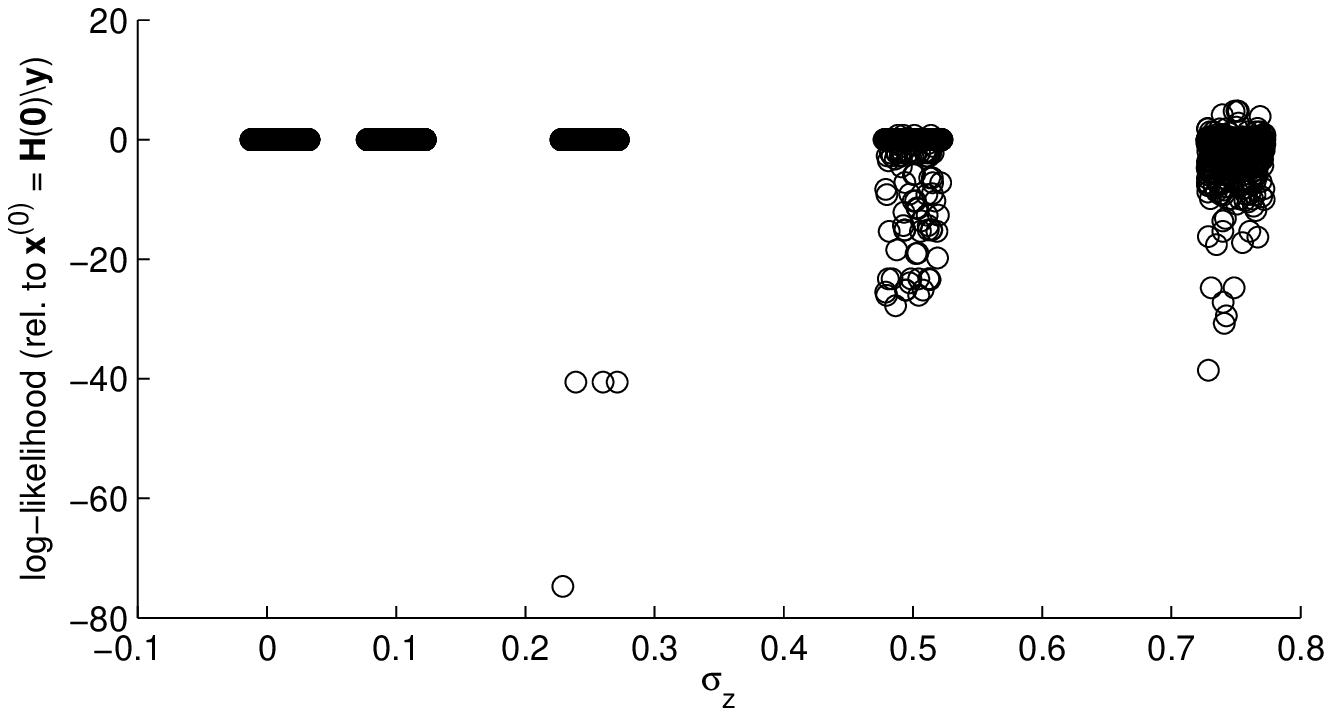}\label{fig:initscat-sigma_zs}}\\
\subfloat[][$K = 10$, $M = 8$, $\sigma_z = 0.25$, and varying $\sigma_w$.]{\includegraphics[width=3.45in]{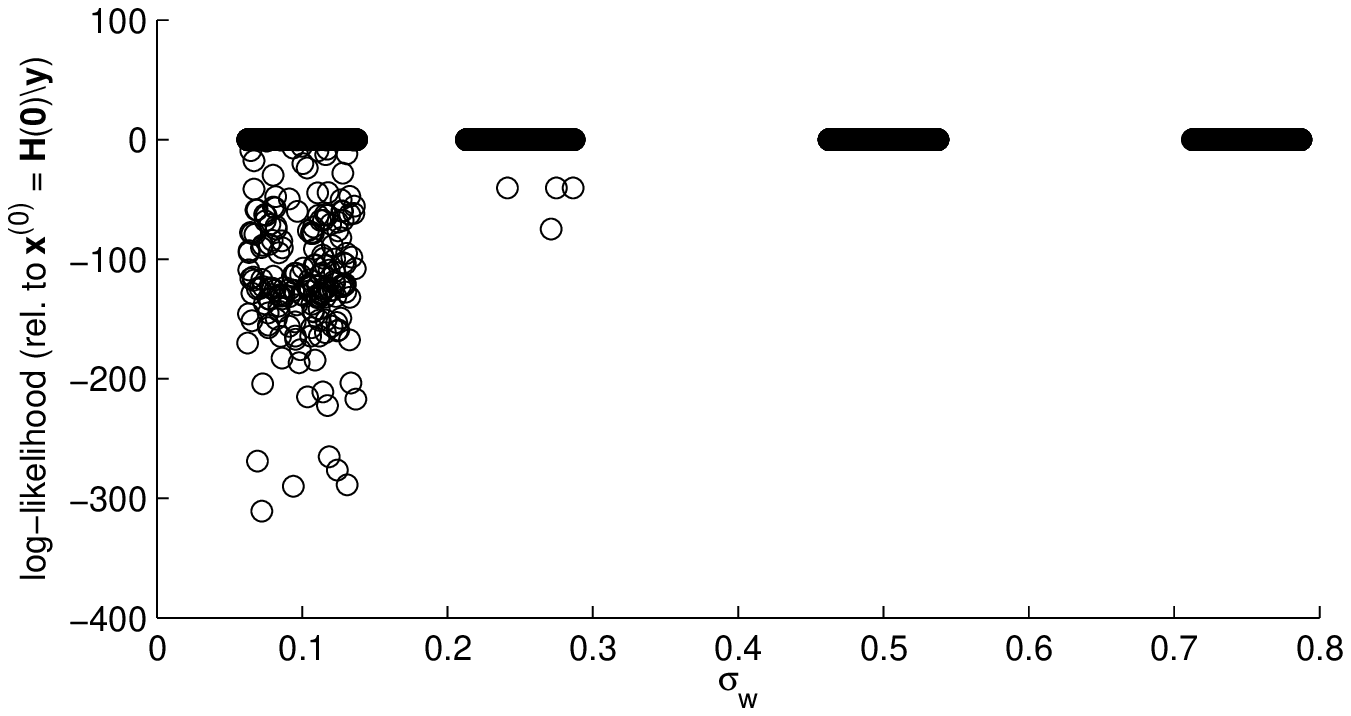}\label{fig:initscat-sigma_ws}}
\caption{The effects of varying initial conditions of the EM algorithm as a function of \protect\subref{fig:initscat-Ms} oversampling factor, \protect\subref{fig:initscat-sigma_zs} jitter variance, and \protect\subref{fig:initscat-sigma_ws} additive noise variance are studied by computing the log-likelihoods of the EM algorithm results, for multiple initial conditions, across $50$ trials. The log-likelihood of the EM algorithm results are displayed relative to the result for zero-jitter initialization, so that the log-likelihood of the result for $\mathbf{\hat{x}}^{(0)} = \mathbf{H(0)}^\dag\mathbf{y}$ is zero.\label{fig:initscat}}
\end{figure}

The likelihood function described in~\eqref{eq:intro_lfunc} is generally nonconcave, so maximizing the function via a hill-climbing method like the EM algorithm is only guaranteed to yield a local maximum. The ability of the algorithm to converge to the global maximum depends on the nonconcavity of the likelihood function. To demonstrate the sensitivity of the EM algorithm, as a function of $M$, $\sigma_z$, and $\sigma_w$, the empirical distribution of the log-likelihood of the optimal values reached from multiple initial conditions is evaluated over numerous trials for different choices of these parameters. In this experiment, the true value of $\mathbf{x}$, the no-jitter linear estimator~\eqref{eq:lin_blueestnoz}, $\mathbf{x = 0}$, and ten random choices, are used as initial conditions for each trial. As suggested by the spread of the samples shown in Figure~\ref{fig:initscat}, the variability of the EM algorithm increases with $\sigma_z$ and decreasing $\sigma_w$. Even when the EM algorithm appears sensitive to initial conditions, using the no-jitter linear estimate~\eqref{eq:lin_blueestnoz} results in a relatively small deviation from the best observed log-likelihood value. In situations when such initialization does fail to produce consistent results, methods such as the deterministic annealing EM algorithm described in~\cite{Ueda98} may improve consistency.

\subsection{Performance of the EM Algorithm}

\begin{figure}
\centering
\subfloat[][]{\includegraphics[width=3.45in]{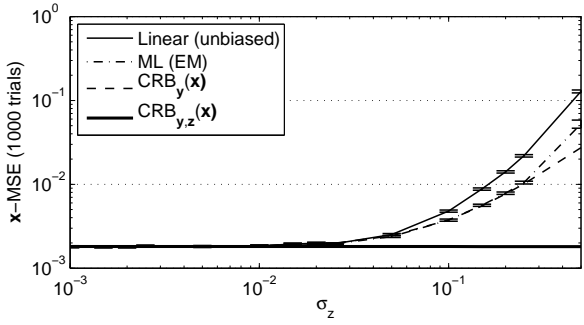}\label{fig:crb-1}}\\
\subfloat[][]{\includegraphics[width=3.45in]{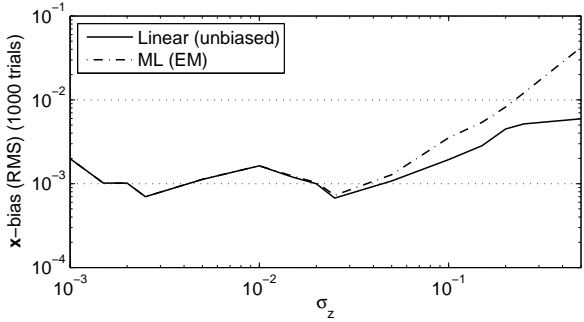}\label{fig:crb-1-bias}}
\caption{The approximate performances of the linear unbiased estimator and ML estimator (EM algorithm) are plotted in \protect\subref{fig:crb-1} against the $\text{CRB}_{\mathbf{y}}(\mathbf{x})$ and the complete data $\text{CRB}_{\mathbf{y,z}}(\mathbf{x})$, as a function of $\sigma_z$ for $K = 10$, $M = 16$, $\sigma_w = 0.05$, and a fixed random choice of $\mathbf{x}$. The linear and ML estimator biases are plotted in \protect\subref{fig:crb-1-bias}, using the root-mean-squared (RMS) values of the bias vectors. The bars above and below each data point for the linear and ML estimators in \protect\subref{fig:crb-1} delineate the $95\%$ confidence intervals for those data points.\label{fig:crb}}
\end{figure}

In the first performance experiment, the Cram\'{e}r--Rao lower bound is compared to the unbiased linear estimator~\eqref{eq:lin_linestpinv} and the EM algorithm of the ML estimator to measure the efficiency of the algorithms. The Cram\'{e}r--Rao lower bound for the complete data is also presented for reference. Although computational difficulties prevent a complete comparison for every possible value of $\mathbf{x}$, carrying out a comparison for a few randomly chosen values of $\mathbf{x}$ provide a measure of the quality of the algorithms. As the curves in Figure~\ref{fig:crb} demonstrate for one such random choice of $\mathbf{x}$, both algorithms are approximately efficient for small $\sigma_{z}$, but the EM algorithm continues to be efficient for larger values of $\sigma_{z}$ than the linear estimator. In addition, the bias shown for the linear and ML estimators is approximately the same for $\sigma_z < \sigma_w$; this bias may be due to the small error in the numerical integration. Note how this error becomes larger with $\sigma_z$.

\begin{figure}
\centering
\subfloat[][$K = 10$, $M = 4$, $0.01 \leq \sigma_z \leq 0.5$, $\sigma_w = 0.05$.]{\includegraphics[width=3.45in]{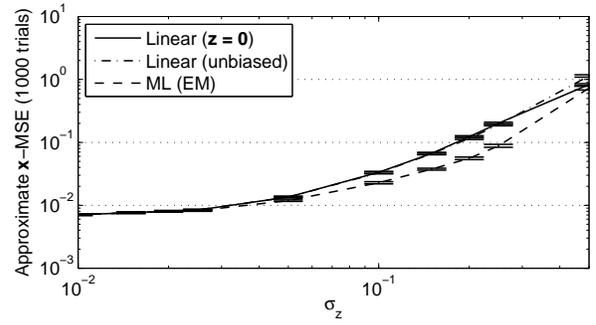}\label{fig:perf-1}}\\
\subfloat[][$K = 10$, $M = 16$, $0.01 \leq \sigma_z \leq 0.5$, $\sigma_w = 0.05$.]{\includegraphics[width=3.45in]{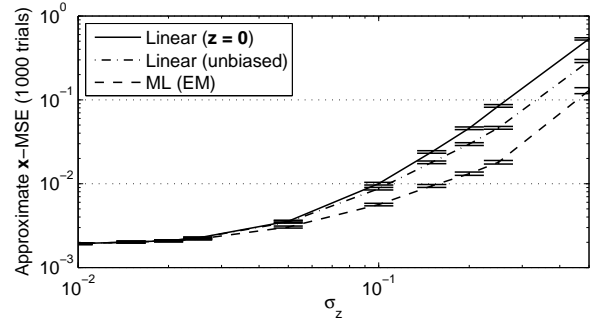}\label{fig:perf-2}}\\
\subfloat[][$K = 10$, $M = 16$, $0.01 \leq \sigma_z \leq 0.5$, $\sigma_w = 0.01$.]{\includegraphics[width=3.45in]{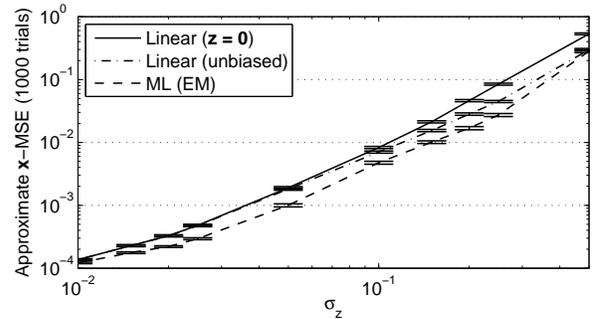}\label{fig:perf-3}}\\
\caption{The MSE performance of the ML estimator (EM algorithm) is compared against both the unbiased linear estimator~\eqref{eq:lin_linestpinv} and the no-jitter BLUE~\eqref{eq:lin_blueestnoz}, as a function of $\sigma_z$. The bars above and below each data point for the linear and ML estimators delineate the $95\%$ confidence intervals for those data points.\label{fig:perf}}
\end{figure}

In Figure~\ref{fig:perf}, the EM algorithm is compared against two linear estimators. First, to demonstrate the MSE improvement attainable through nonlinear estimation, the EM algorithm is pitted against the linear unbiased estimator. Since a major motivating factor for developing these algorithms is to reduce the power consumption due to clock accuracy, the EM algorithm also can achieve the same MSE as the linear estimator for a substantially larger jitter variance, reducing the clock's power consumption.

When the additive noise dominates the jitter ($\sigma_{z} \ll \sigma_{w}$), the improvement can be expected to be minimal, since the system is nearly linear, and the jitter is statistically insignificant. As the amount of jitter increases, the density function $p(\mathbf{z}\mid \mathbf{y};\mathbf{x})$ used in each iteration of the EM algorithm becomes more nonlinear in $\mathbf{z}$, and the quadrature becomes less accurate for a given number of terms. Therefore, the EM algorithm generally takes longer to converge, and the result should be a less accurate approximation to the true ML estimator. This behavior is observed in Figure~\ref{fig:perf}, where the EM algorithm is compared against both the linear unbiased estimator and the no-jitter linear estimator. The EM algorithm generally has lower MSE than either linear estimator, and the performance gap is more pronounced for higher oversampling factor $M$.

\begin{figure}
\centering
\subfloat[][$K = 10$, $M$ varies, $0.01 \leq \sigma_z \leq 0.5$, $\sigma_w = 0.1$.]{\includegraphics[width=3.45in]{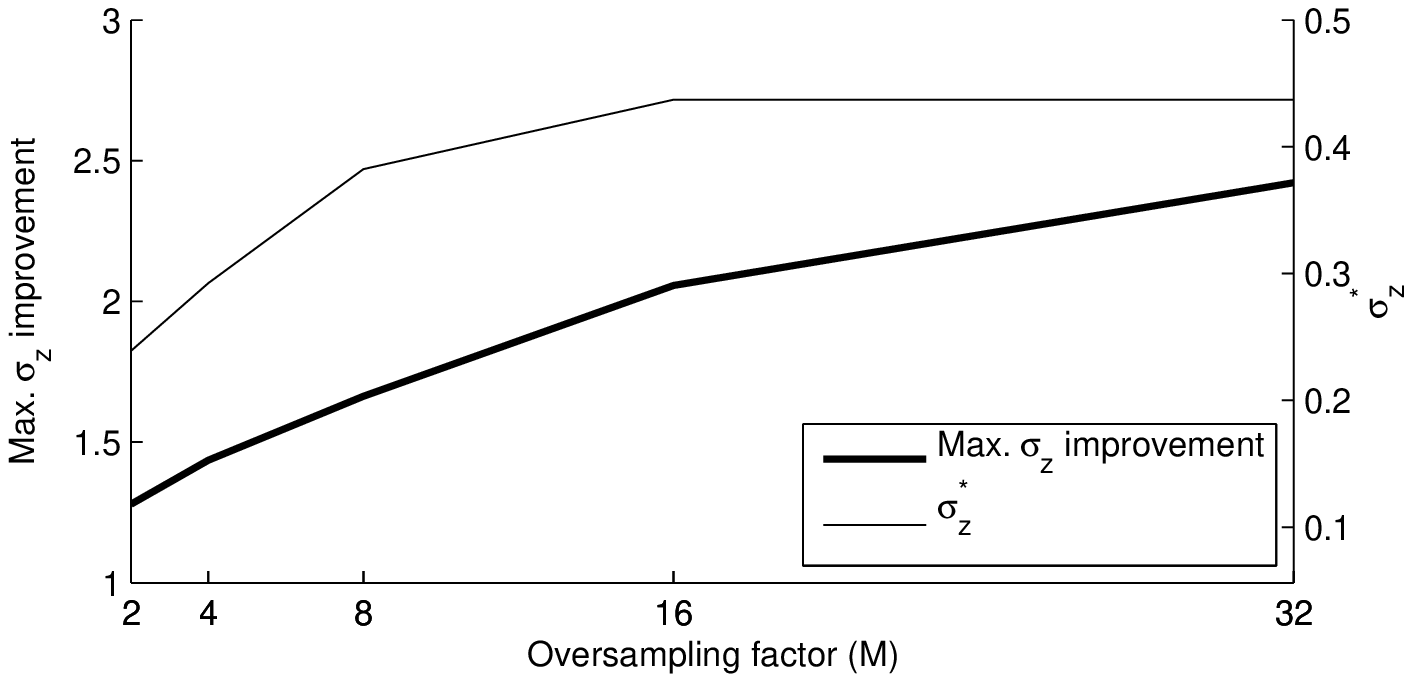}\label{fig:jitterimprove-Ms}}\\
\subfloat[][$K = 10$, $M = 8$, $0.01 \leq \sigma_z \leq 0.5$, $\sigma_w$ varies.]{\includegraphics[width=3.45in]{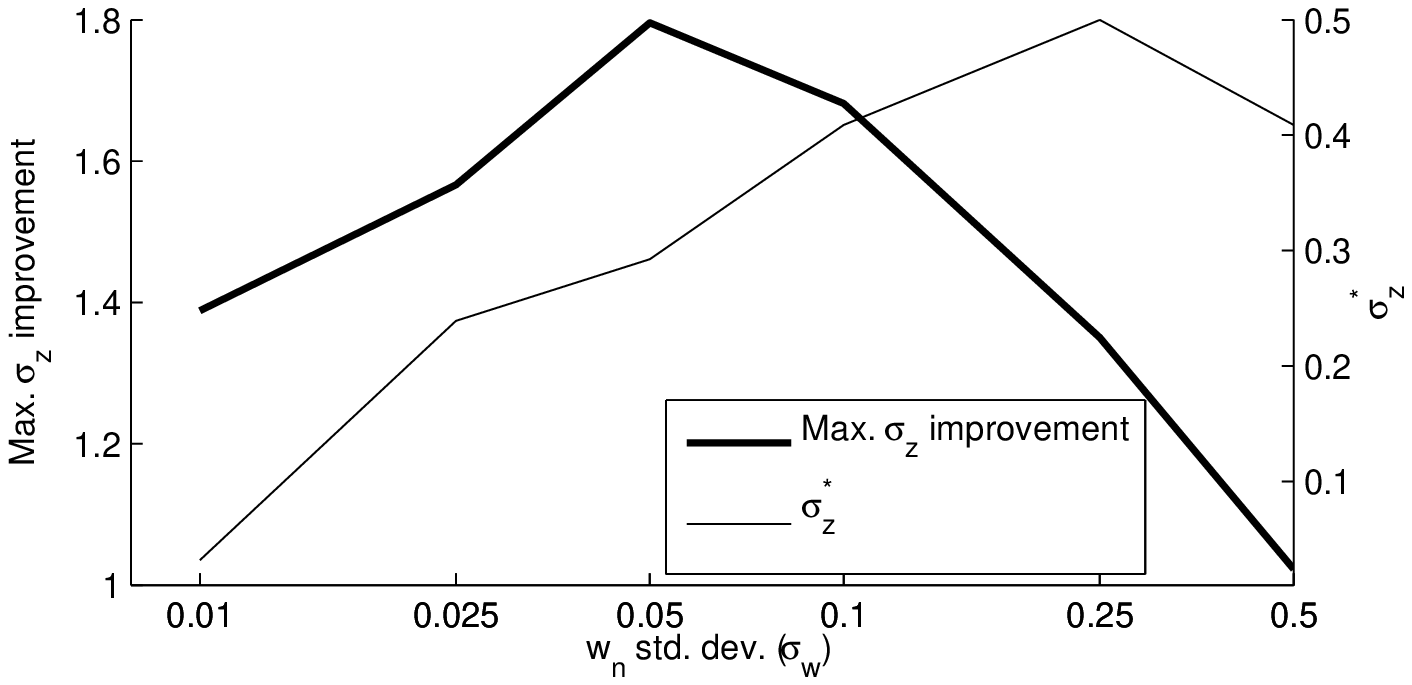}\label{fig:jitterimprove-sigmaws}}\\
\caption{These graphs show the maximum factor of improvement in jitter tolerance, measured by $\sigma_z$, achievable by the EM algorithm (relative to linear reconstruction). Holding $\sigma_w$ fixed, \protect\subref{fig:jitterimprove-Ms} shows the trend in maximum improvement as $M$ increases, and \protect\subref{fig:jitterimprove-sigmaws} shows the trend in maximum improvement as $\sigma_w$ increases while holding $M$ fixed. The jitter standard deviation $\sigma_z^*$ corresponding to this maximum improvement for the ML estimator is plotted on the same axes.\label{fig:jitterimproves}}
\end{figure}

To answer the question of how much more jitter can be tolerated for the same desired MSE using the EM algorithm, the maximum proportional increase is plotted as a function of $M$ and $\sigma_{w}$ in Figure~\ref{fig:jitterimproves}. The maximum proportional increase for a choice of $M$ and $\sigma_{w}$ is computed by approximating log-log domain MSE curves, like those in Figure~\ref{fig:perf}, with piece-wise linear curves and interpolating the maximum distance between them over the range of $\sigma_z \geq \sigma_w$. The range of $\sigma_z < \sigma_w$ is ignored since the linear and nonlinear reconstructions perform similarly when the additive noise dominates (as expected). The proportion of improvement increases linearly as $M$ increases. As $\sigma_w$ increases, the level of improvement stays approximately the same for $\sigma_w < \sigma_z$. However, when $\sigma_w$ increases beyond $\sigma_z$, the level of improvement decreases substantially as expected, since the additive noise dominates, and the optimal estimator is approximately linear. A maximum $\sigma_z$ improvement factor of two corresponds to power savings of up to $75$ percent.

%Simulations considering proposed EM algorithm, and comparisons to the other estimators, and the CRB.
%
%\subsection{Convergence Analysis}
%Include some of the simulations from Section 3.5.1 of the thesis.
%
%\subsection{Performance Comparisons}
%Compare MSE (or SNR) performance of EM algorithm against linear, iterative/recursive, and MCMC methods. Also, compute the CRB and compare EM algorithm.
%

% conclusions, further work
\section{Conclusion}\label{sec:conclusion}

The results presented in Section~\ref{sec:simresults} are very encouraging from a power-consumption standpoint. A maximum improvement of between $1.4$ to $2$ times the jitter translates to a two-to-fourfold decrease in power consumption by the clock, according to~\eqref{eq:intro_SAPtradeoff}. To put the magnitude of such an improvement in context, consider the digital baseband processor for ultra-wideband communication in~\cite{Blazquez05}. This processor incorporates an ADC and a PLL, which consume $86$ mW and $45$ mW, respectively, out of a $271$ mW budget for the chip. Reducing by a factor of two the power consumed by the ADC alone would decrease the total power consumption of the chip by almost sixteen percent. 

While effective, the EM algorithm is computationally expensive. One benefit of digital post-processing is that these algorithms can be performed off-chip, on a computer or other system with less limited computational resources. For real-time on-chip applications, Kalman filter-like versions of the EM algorithm would be more practical; this extension is a topic for further investigation. Related to real-time processing is developing streaming algorithms for the infinite-dimensional case, extending this work for general real-time sampling systems. Another future direction involves modifying these algorithms for correlated or periodic jitter.

Sampling jitter mitigation is actually just one application of these new algorithms. In the frequency domain, jitter maps to uncertainty in frequency; using algorithms such as these should produce more reliable Fourier transforms for systems like spectrum analyzers. In higher dimensions, timing noise becomes location jitter in images or video. Greater tolerance of the locations of pixels in images would allow scanning electron microscope users to acquire higher resolution images without sacrificing MSE. This paper shows that significant improvements over the best linear post-processing are possible; thus, further work may impact these and other applications.

% if have a single appendix:
%\appendix[Proof of the Zonklar Equations]
% or
%\appendix  % for no appendix heading
% do not use \section anymore after \appendix, only \section*
% is possibly needed

% use appendices with more than one appendix
% then use \section to start each appendix
% you must declare a \section before using any
% \subsection or using \label (\appendices by itself
% starts a section numbered zero.)
%

%\appendices
%\input{appderiveblue}
%\section{Proof of the First Zonklar Equation}
%Appendix one text goes here.
%
% you can choose not to have a title for an appendix
% if you want by leaving the argument blank
%\section{}
%Appendix two text goes here.
%
%
% use section* for acknowledgement
\section*{Acknowledgment}

The authors thank J. Kusuma for asking stimulating questions about sampling and applications of jitter mitigation. The authors also thank Z. Zvonar at Analog Devices and G. Frantz at Texas Instruments for their insights and support.

% Can use something like this to put references on a page
% by themselves when using endfloat and the captionsoff option.
\ifCLASSOPTIONcaptionsoff
  \newpage
\fi

% trigger a \newpage just before the given reference
% number - used to balance the columns on the last page
% adjust value as needed - may need to be readjusted if
% the document is modified later
\ifCLASSOPTIONdraftcls
\else
	\IEEEtriggeratref{22}
\fi
% The "triggered" command can be changed if desired:
%\IEEEtriggercmd{\enlargethispage{-5in}}

% references section

% can use a bibliography generated by BibTeX as a .bbl file
% BibTeX documentation can be easily obtained at:
% http://www.ctan.org/tex-archive/biblio/bibtex/contrib/doc/
% The IEEEtran BibTeX style support page is at:
% http://www.michaelshell.org/tex/ieeetran/bibtex/
\bibliographystyle{IEEEtran}
% argument is your BibTeX string definitions and bibliography database(s)
\bibliography{IEEEabrv,TSP-jittercomp-classical}

% Generated by IEEEtran.bst, version: 1.12 (2007/01/11)
\begin{thebibliography}{10}
\providecommand{\url}[1]{#1}
\csname url@samestyle\endcsname
\providecommand{\newblock}{\relax}
\providecommand{\bibinfo}[2]{#2}
\providecommand{\BIBentrySTDinterwordspacing}{\spaceskip=0pt\relax}
\providecommand{\BIBentryALTinterwordstretchfactor}{4}
\providecommand{\BIBentryALTinterwordspacing}{\spaceskip=\fontdimen2\font plus
\BIBentryALTinterwordstretchfactor\fontdimen3\font minus
  \fontdimen4\font\relax}
\providecommand{\BIBforeignlanguage}[2]{{%
\expandafter\ifx\csname l@#1\endcsname\relax
\typeout{** WARNING: IEEEtran.bst: No hyphenation pattern has been}%
\typeout{** loaded for the language `#1'. Using the pattern for}%
\typeout{** the default language instead.}%
\else
\language=\csname l@#1\endcsname
\fi
#2}}
\providecommand{\BIBdecl}{\relax}
\BIBdecl

\bibitem{NationalNote103}
I.~King, ``Understanding high-speed signals, clocks, and data capture,''
  National Semicondutor Signal Path Designer, vol. 103, Apr. 2005.

\bibitem{Lee00}
T.~H. Lee and A.~Hajimiri, ``Oscillator phase noise: A tutorial,'' \emph{{IEEE}
  J. Solid-State Circuits}, vol.~35, no.~3, pp. 326--336, Mar. 2000.

\bibitem{Uyttenhove01}
K.~Uyttenhove and M.~S.~J. Steyaert, ``Speed-power-accuracy tradeoff in
  high-speed {CMOS} {ADC}s,'' \emph{{IEEE} Trans. Circuits Syst. {II}},
  vol.~49, no.~4, pp. 280--287, Apr. 2002.

\bibitem{Brannon00}
B.~Brannon, ``Aperture uncertainty and {ADC} system performance,'' Analog
  Devices, Tech. Rep. AN-501, Sep. 2000.

\bibitem{Walden99}
R.~H. Walden, ``Analog-to-digital converter survey and analysis,'' \emph{{IEEE}
  J. Sel. Areas Commun.}, vol.~17, no.~4, pp. 539--550, Apr. 1999.

\bibitem{Lee02}
Y.~Lee and R.~Mittra, ``Electromagnetic interference mitigation by using a
  spread-spectrum approach,'' \emph{{IEEE} Trans. Electromagn. Compat.},
  vol.~44, no.~2, pp. 380--385, May 2002.

\bibitem{Balakrishnan62}
A.~V. Balakrishnan, ``On the problem of time jitter in sampling,'' \emph{{IRE}
  Trans. Inform. Th.}, vol.~8, no.~3, pp. 226--236, Apr. 1962.

\bibitem{Brown63}
W.~M. Brown, ``Sampling with random jitter,'' \emph{J. Soc. Industrial and
  Appl. Math.}, vol.~11, no.~2, pp. 460--473, Jun. 1963.

\bibitem{Liu65}
B.~Liu and T.~P. Stanley, ``Error bounds for jittered sampling,'' \emph{{IEEE}
  Trans. Autom. Control}, vol.~10, no.~4, pp. 449--454, Oct. 1965.

\bibitem{Feichtinger95}
H.~G. Feichtinger, K.~Gr\"{o}chenig, and T.~Strohmer, ``Efficient numerical
  methods in non-uniform sampling theory,'' \emph{Numer. Math.}, vol.~69,
  no.~4, pp. 423--440, Feb. 1995.

\bibitem{Marziliano00}
P.~Marziliano and M.~Vetterli, ``Reconstruction of irregularly sampled
  discrete-time bandlimited signals with unknown sampling locations,''
  \emph{{IEEE} Trans. Signal Process.}, vol.~48, no.~12, pp. 3462--3471, Dec.
  2000.

\bibitem{Tuncer07}
T.~E. Tuncer, ``Block-based methods for the reconstruction of finite-length
  signals from nonuniform samples,'' \emph{{IEEE} Trans. Signal Process.},
  vol.~55, no.~2, pp. 530--541, Feb. 2007.

\bibitem{Divi04}
V.~Divi and G.~Wornell, ``Signal recovery in time-interleaved analog-to-digital
  converters,'' in \emph{Proc. {IEEE} Int. Conf. Acoustics, Speech, Signal
  Process.}, vol.~2, May 2004, pp. 593--596.

\bibitem{WellerThesis}
D.~S. Weller, ``Mitigating timing noise in {ADC}s through digital
  post-processing,'' {SM} Thesis, Massachusetts Institute of Technology,
  Department of Electrical Engineering and Computer Science, Jun. 2008.

\bibitem{Kythe05}
P.~K. Kythe and M.~R. Sch\"{a}ferkotter, \emph{Handbook of Computational
  Methods for Integration}.\hskip 1em plus 0.5em minus 0.4em\relax Boca Raton,
  FL: CRC, 2005.

\bibitem{Golub69}
G.~H. Golub and J.~H. Welsch, ``Calculation of {G}auss quadrature rules,''
  \emph{Math. Comp.}, vol.~23, no. 106, pp. 221--230, Apr. 1969.

\bibitem{Davis84}
P.~J. Davis and P.~Rabinowitz, \emph{Methods of Numerical Integration}.\hskip
  1em plus 0.5em minus 0.4em\relax Orlando: Academic Press, 1984.

\bibitem{Dempster77}
A.~P. Dempster, N.~M. Laird, and D.~B. Rubin, ``Maximum likelihood from
  incomplete data via the {EM} algorithm,'' \emph{J. Roy. Statist. Soc., Ser.
  B}, vol.~39, no.~1, pp. 1--38, 1977.

\bibitem{Herzet07}
C.~Herzet and L.~Vandendorpe, ``Prediction of the {EM}-algorithm speed of
  convergence with {C}ramer-{R}ao bounds,'' in \emph{Proc. {IEEE} Int. Conf.
  Acoustics, Speech, Signal Process.}, vol.~3, Apr. 2007, pp. 805--808.

\bibitem{Meng91}
X.-L. Meng and D.~B. Rubin, ``Using {EM} to obtain asymptotic
  variance-covariance matrices: The {SEM} algorithm,'' \emph{J. Amer. Statist.
  Assoc.}, vol.~86, no. 416, pp. 899--909, Dec. 1991.

\bibitem{Kirkpatrick83}
S.~Kirkpatrick, C.~D. Gelatt, and M.~P. Vecchi, ``Optimization by simulated
  annealing,'' \emph{Science}, vol. 220, no. 4598, pp. 671--680, May 1983.

\bibitem{Kay93}
S.~M. Kay, \emph{Fundamentals of Statistical Signal Processing: Estimation
  Theory}, ser. Prentice-Hall Signal Processing Series.\hskip 1em plus 0.5em
  minus 0.4em\relax Upper Saddle River, NJ: Prentice Hall, 1993, vol.~1.

\bibitem{Kusuma08}
J.~Kusuma and V.~K. Goyal, ``Delay estimation in the presence of timing
  noise,'' \emph{{IEEE} Trans. Circuits Syst. {II}}, vol.~55, no.~9, pp.
  848--852, Sep. 2008.

\bibitem{Ueda98}
N.~Ueda and R.~Nakano, ``Deterministic annealing {EM} algorithm,'' \emph{Neural
  Networks}, vol.~11, no.~2, pp. 271--282, Mar. 1998.

\bibitem{Blazquez05}
R.~Bl\'{a}zquez, P.~P. Newaskar, F.~S. Lee, and A.~P. Chandrakasan, ``A
  baseband processor for impulse ultra-wideband communications,'' \emph{{IEEE}
  J. Solid-State Circuits}, vol.~40, no.~9, pp. 1821--1828, Sep. 2005.

\end{thebibliography}
\end{document}